\documentclass[english]{article}
\usepackage[T1]{fontenc}
\usepackage[a4paper]{geometry}
\usepackage{amsmath}
\usepackage{graphicx}
\usepackage{setspace}
\usepackage{esint}
\usepackage{url}
\usepackage{color}

\usepackage{cite}

\usepackage[normalem]{ulem}
\definecolor{pkcolor}{rgb}{0,0,1}

\newcommand\pkout{\bgroup\markoverwith{\color{pkcolor}{\rule[0.4ex]{2pt}{0.8pt}}}\ULon}

\usepackage[normalem]{ulem}
\definecolor{avhcolor}{rgb}{1,0,0}

\newcommand\avhout{\bgroup\markoverwith{\color{avhcolor}{\rule[0.4ex]{2pt}{0.8pt}}}\ULon}

\newcommand\vhlib{{\sc A Very Handy LIBrary }}
\newcommand\ugdKS{\texttt{KS}}
\newcommand\ugdKSSud{\texttt{KS+Sudakov}}
\newcommand\ugdKSmu{\texttt{Kutak-nonlinear-PRD15}}
\newcommand\ugdKMR{\texttt{KMR}}
\newcommand\ugdLHC{\texttt{LHC-fit}}
\newcommand\ugdLHCSud{\texttt{LHC-fit+Sudakov}}

\usepackage{babel}

\begin{document}

\title{%
\hspace{\fill}{\normalsize IFJPAN-IV-2015-8}\\[2ex]
Resummation effects in forward production of \\ $Z_{0}$+jet at LHC%
}

\author{A.~van~Hameren, P.~Kotko, K.~Kutak}

\author{
A.~van Hameren,$^1$ P.~Kotko,$^2$ K.~Kutak,$^1$ \\\\
$^1$ {\small\it The H.\ Niewodnicza\'nski Institute of Nuclear Physics PAN}\\ {\small\it Radzikowskiego 152, 31-342 Krak\'ow, Poland}\\\\
$^2$ {\small\it Department of Physics, The Pennsylvania State University}\\{\small\it University Park, PA 16802, United States,}\\
}

\date{}

\maketitle

\begin{abstract}
We calculate several differential
cross sections for $Z_{0}$ and high-$p_{T}$ jet production in the
forward rapidity region at the LHC using the hybrid High Energy Factorization.
We test various unintegrated gluon
distributions involving subleading BFKL effects (such as kinematic constraint,
running strong coupling and DGLAP correction) and compare the results
with experimental data obtained by the LHCb experiment. 
We find that
the hard scale dependence of unintegrated gluon distributions, which effectively resums the Sudakov-type logarithms on the top of the resummation of the small $x$ logarithms, is essential
to describe the normalized azimuthal decorrelations between the $Z_0$-boson and the jet.
\end{abstract}

\section{Motivation}
\label{sec:intro}

The Large Hadron Collider 
 opens an opportunity to explore kinematic
regions where particles produced in high-energy collisions posses
large transverse momenta and rapidities. The production of electroweak bosons and jets is a vital part of tests of Standard Model as well as 
searches of physics beyond the Standard Model. Furthermore it has been recognized in \cite{Dooling:2014kia} that by studies of associated production of electroweak bosons and jets 
may provide a new insight into the 
transverse  partonic structure of hadrons at small
$x$, where $x$ is the momentum fraction of the hadron taken by a parton
participating in the hard collision.
Furthermore, such a final state, being a combination of colorful and colorless particles, gives the opportunity for particularly interesting investigations complementary to results obtained in studies of pure jet final states \cite{Deak:2010gk,vanHameren:2013fla,Nefedov:2013ywa,vanHameren:2014ala,vanHameren:2014lna} and Drell-Yan pairs \cite{Lipatov:2011sd}.
In particular the final state rescatterings due to soft color exchanges should have less impact on the properties of produced final state as compared to pure jet final states.

This work is motivated by a recent LHCb measurement \cite{Aaij:2013nxa}
at $\sqrt{s}=7\,\mathrm{TeV}$ of the process 
\begin{equation}
pp\rightarrow Z_{0}\left(\rightarrow\mu^{+}\mu^{-}\right)+\textrm{jet}
\end{equation}
in the forward direction within the pseudorapidity range
$2.0<\eta<4.5$. 
The final-state muon and anti-muon were required to have transverse
momenta $p_{T\mu}>20\,\mathrm{GeV}$ while the leading jet was considered
with two different cuts: $p_{T\, j}>10\,\mathrm{GeV}$ and $p_{T\, j}>20\,\mathrm{GeV}$.
The jets were reconstructed using the anti-$k_{T}$ algorithm with radius
$R=0.5$ and they were required to be separated from muon tracks on
the $\phi-\eta$ plane (azimuthal angle-pesudorapidity) by a distance
$R=0.4$. The muon-pair was required to have an invariant mass within the range $60\,\mathrm{GeV<M_{\mu\mu}<120\,}\mathrm{GeV}$.
The rapidity constraint assures that in the partonic picture of the
process, one of the initial state partons carries a rather small fraction
$x_{A}$ of the corresponding hadron momentum $p_{A}$, while the
other must have a fraction $x_{B}\gg x_{A}$ (c.f. Fig.~\ref{fig:diagram}).

In order to describe the process perturbatively, one definitely needs to go beyond the pure collinear factorization and support the calculation by a resummations. In the modern advanced approaches this is achieved by parton showers and hadronization as implemented for example in {\sc Pythia} \cite{Sjostrand:2006za}. In the present work we consider another approach, namely a  resummation of logarithms of $\mathrm{ln}(1/x_A)$ and $\mathrm{ln}(\mu /k_T)$, where $\mu$ is a hard scale and $k_T$ is a certain additional scale given by the imbalance of the final states on the transverse plane. This approach captures certain aspects of the process more accurately already at lowest order of the strong coupling constant in the hard process.

In the present paper, we will therefore attempt to study the process within so called High Energy Factorization, or more precisely, using so-called hybrid High Energy Factorization 
motivated by the works
\cite{Catani:1990eg,Catani:1990xk,Catani:1994sq,Deak:2009xt}.
Within the asymmetric kinematic situation $x_{B}\gg x_{A}$ described above, the cross section for the process under consideration can be expressed by the following formula
\begin{multline}
d\sigma_{AB\rightarrow\mu^{+}\mu^{-}+\mathrm{jet}+X}=\int d^{2}k_{TA}\int\frac{dx_{A}}{x_{A}}\,\int dx_{B}\,\sum_{b}\\
\times\mathcal{F}_{g^{*}/A}\left(x_{A},k_{T\, A},\mu\right)\, f_{b}\left(x_{B},\mu\right)\, d\hat{\sigma}_{g^{*}q_b\rightarrow q_b\mu^{+}\mu^{-}}\left(x_{A},x_{B},k_{T\, A},\mu\right),\label{eq:HEN_fact_2}
\end{multline}
where $\mathcal{F}_{g^{*}/A}$ is the unintegrated gluon distribution 
for hadron $A$, $f_{b}$ is a collinear PDF and $d\hat{\sigma}_{g^{*}q_b\rightarrow q_b\mu^{+}\mu^{-}}$
is the hard cross section obtained from a gauge invariant tree-level
off-shell amplitude for the process $g^{*}q_b\rightarrow q_b\mu^{+}\mu^{-}$, and where $q_b$ refers to quarks as well as anti-quarks
(Fig.~\ref{fig:diagram}). 

Let us note, that the original High Energy Factorization prescription was designed to study inclusive small $x$ processes and the corresponding unintegrated gluon distribution was assumed to undergo the Balitski-Fadin-Kuraev-Lipatov (BFKL) evolution equation
(see e.g. \cite{Lipatov:1996ts}). For more exclusive processes it is however necessary to include the subleading BFKL effects, such as kinematic constraint ensuring energy conservation, large-$x$ correction, running strong coupling constant, and -- notably -- the hard scale dependence. The last, denoted in Eq.~(\ref{eq:HEN_fact_2}) as $\mu$, turns out to be essential to describe the data under consideration.
In the present work we shall not discuss the validity of the model (\ref{eq:HEN_fact_2}) on the theoretical level. Instead, we shall test it phenomenologically against the existing data.
For a more detailed review of various approaches to the small $x$ factorization and related issues see e.g. \cite{Kotko:2015ksa}. For a derivation of hybrid High Energy Factorization from the dilute limit of the Color Glass Condensate approach see \cite{Kotko:2015ura}.

\section{Results}
\label{sec:results}

\begin{figure}
\begin{centering}
\includegraphics[width=0.6\textwidth]{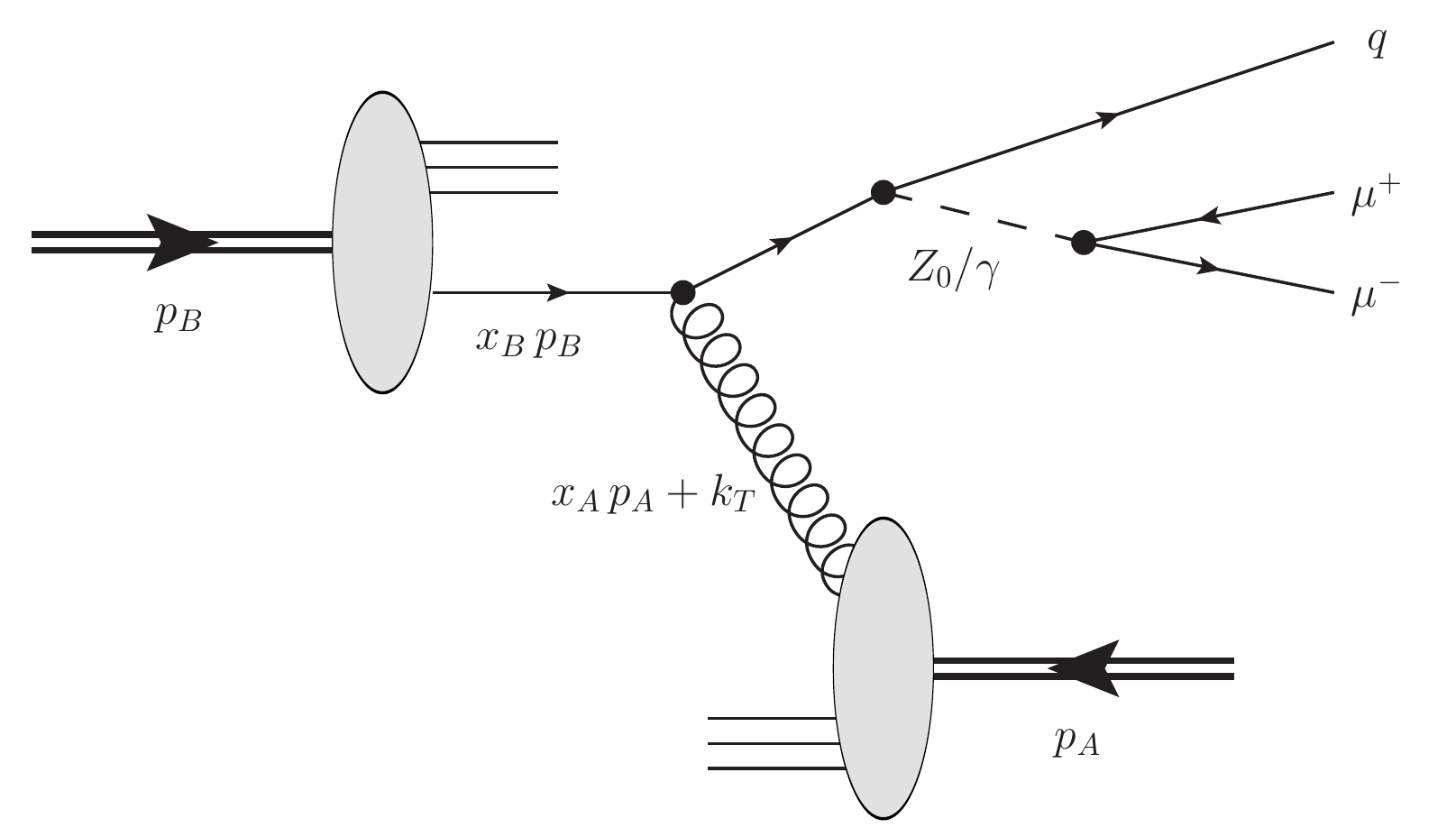}
\par\end{centering}

\caption{\small Diagrammatic representation of the hybrid high energy factorization
for forward $Z_{0}$+jet production. The upper blob corresponds to
a collinear PDF, whereas the lower one corresponds to the unintegrated
gluon distribution. The gluon entering the hard scattering is off-shell
with virtuality $k_{T}^{2}$. \label{fig:diagram}}
\end{figure}

Using the formalism described in the preceding section we have computed the cross sections for $Z_0+\mathrm{jet}$ production.
We have used two programs to calculate off-shell $qg^{*}\rightarrow q \mu^{+} \mu^{-}$ amplitude (with $Z_0$ and $\gamma$ exchange) and to cross-check the results. The first program is \vhlib\ (AVHLIB)~\cite{Bury:2015dla} in {\sc Fortran}, in which the approach of \cite{vanHameren:2012if} is implemented. It computes amplitudes entirely numerically and includes
a full Monte Carlo program.
The second is the electroweak extension of the program {\sc OGIME} \cite{Kotko_OGIME}. It calculates amplitudes analytically
in a form that can be interfaced with a Monte Carlo program. More specifically,
the analytic expressions were implemented in the {\sc C++} code {\sc LxJet} \cite{Kotko_LxJet}.
Note, however, that since there are no final state gluons, the ordinary
Feynman diagram depicted in Fig.~\ref{fig:diagram} (with appropriate
high energy projector for the off-shell gluon as described in \cite{Catani:1990eg})
is enough to obtain the gauge invariant amplitude for this process,
and all the complications discussed in \cite{vanHameren:2012uj,vanHameren:2012if,vanHameren:2013csa,Kotko:2014aba,vanHameren:2014iua}
do not have any impact here.

\begin{figure}
\begin{centering}
\includegraphics{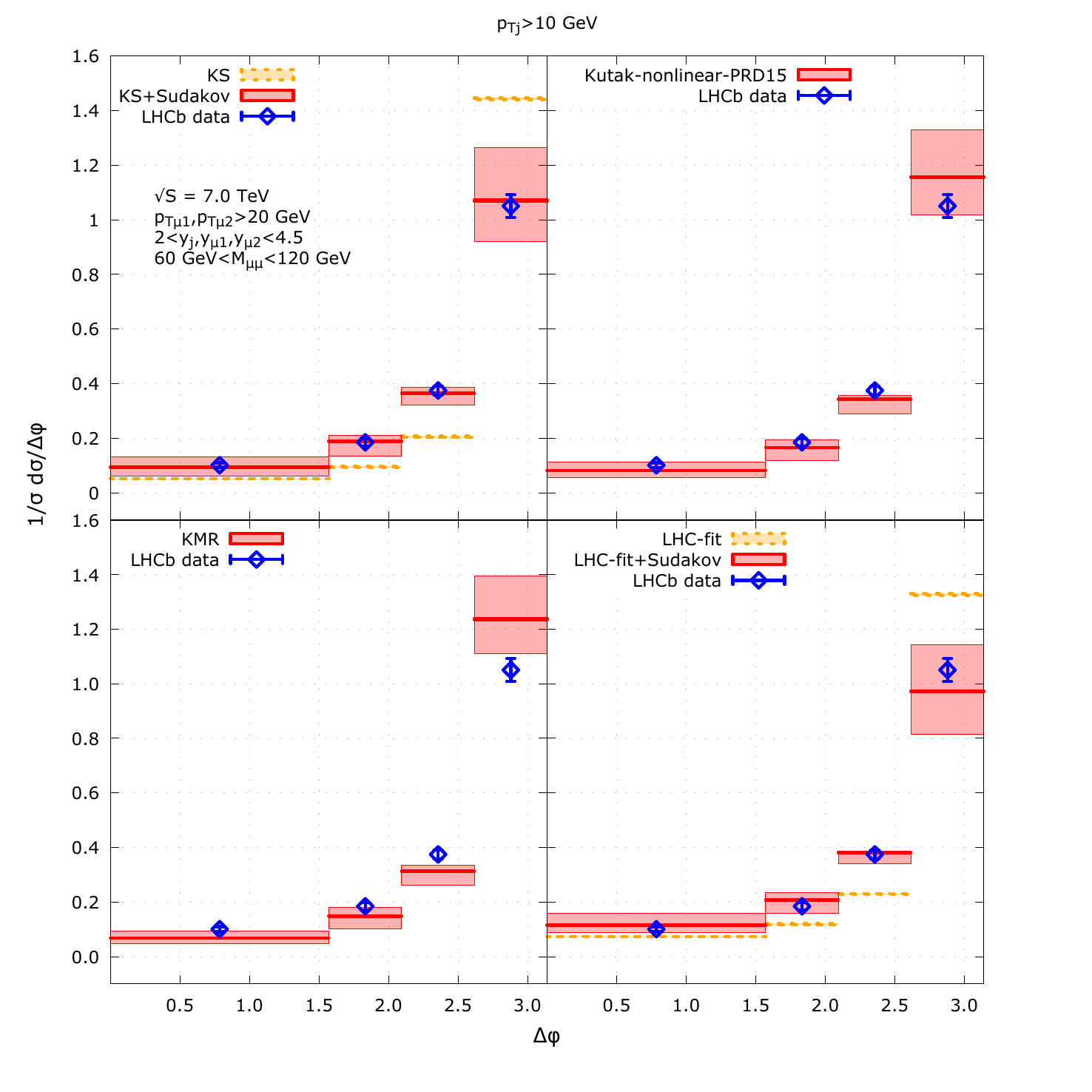}
\par\end{centering}

\caption{\small Azimuthal decorrelations for $p_{T\, j}>10\,\mathrm{GeV}$ normalized
to the total cross section.\label{fig:dphi10}}

\end{figure}

\begin{figure}
\begin{centering}
\includegraphics{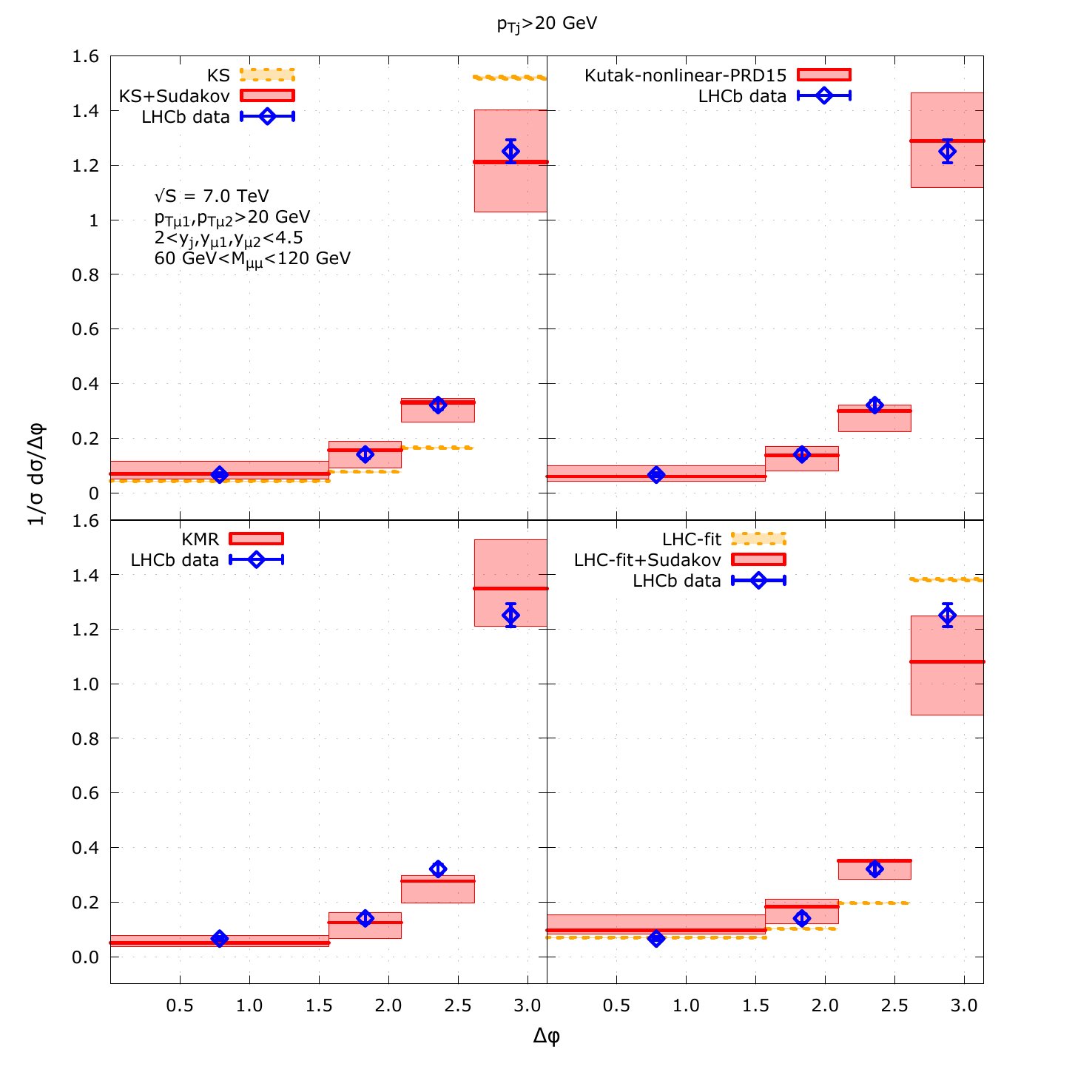}
\par\end{centering}

\caption{\small Azimuthal decorrelations for $p_{T\, j}>20\,\mathrm{GeV}$ normalized
to the total cross section.\label{fig:dphi20}}
\end{figure}

\begin{figure}
\begin{centering}
\includegraphics{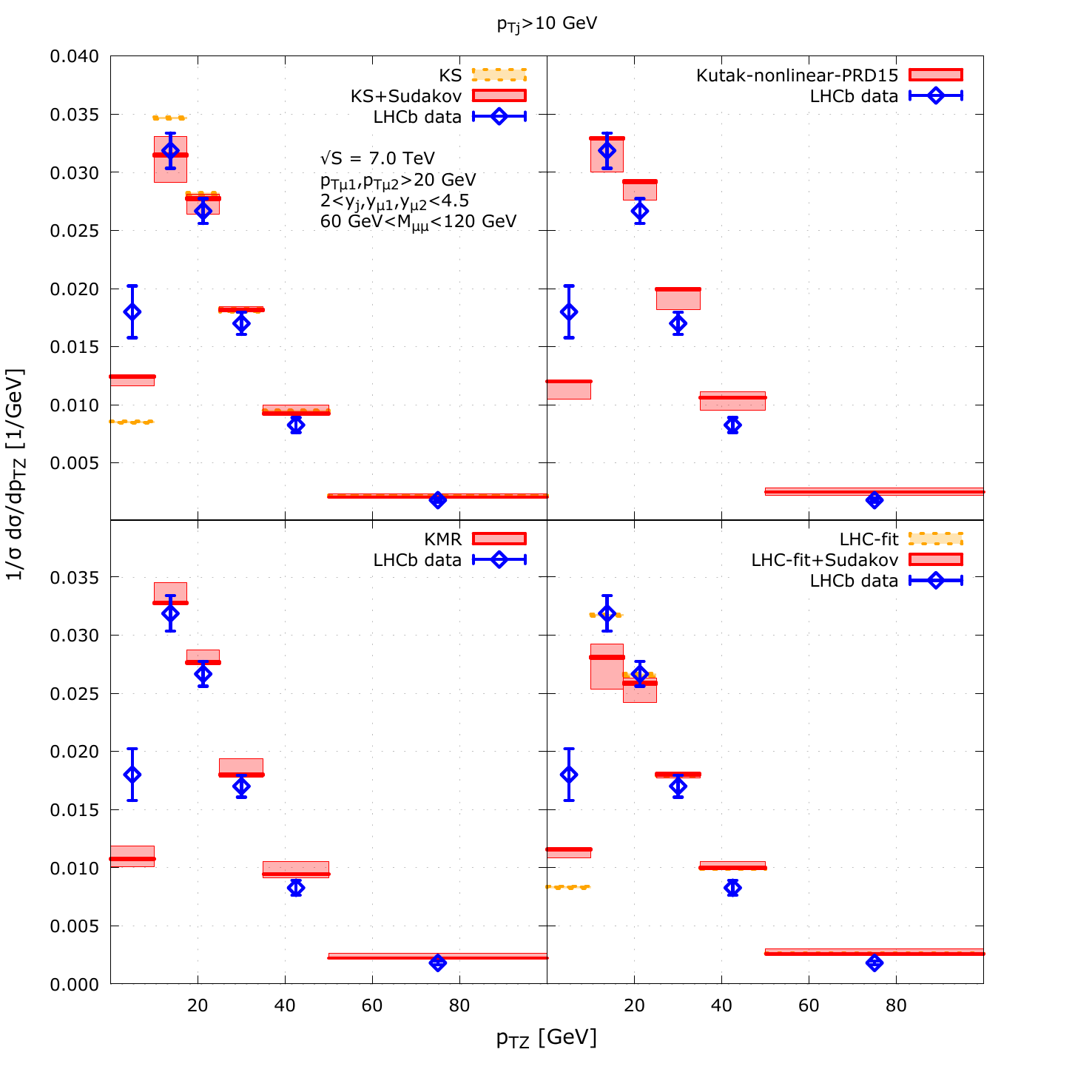}
\par\end{centering}

\caption{\small Transverse momentum spectrum for $Z_{0}$-boson for $p_{T\, j}>10\,\mathrm{GeV}$
normalized to the total cross section.\label{fig:pTZ10}}
\end{figure}

\begin{figure}
\begin{centering}
\includegraphics{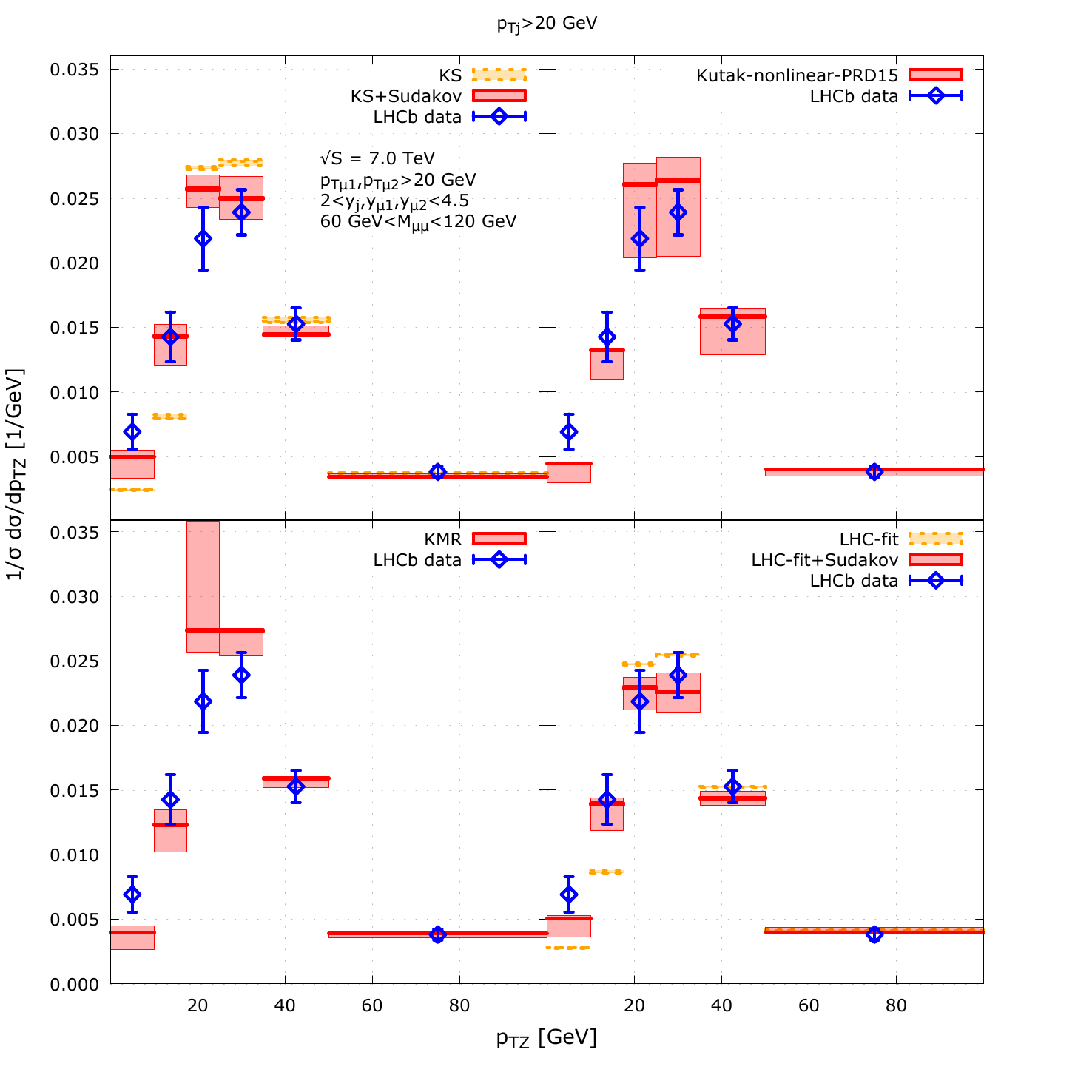}
\par\end{centering}

\caption{\small Transverse momentum spectrum for $Z_{0}$-boson for $p_{T\, j}>20\,\mathrm{GeV}$
normalized to the total cross section.\label{fig:pTZ20}}
\end{figure}

\begin{figure}
\begin{centering}
\includegraphics{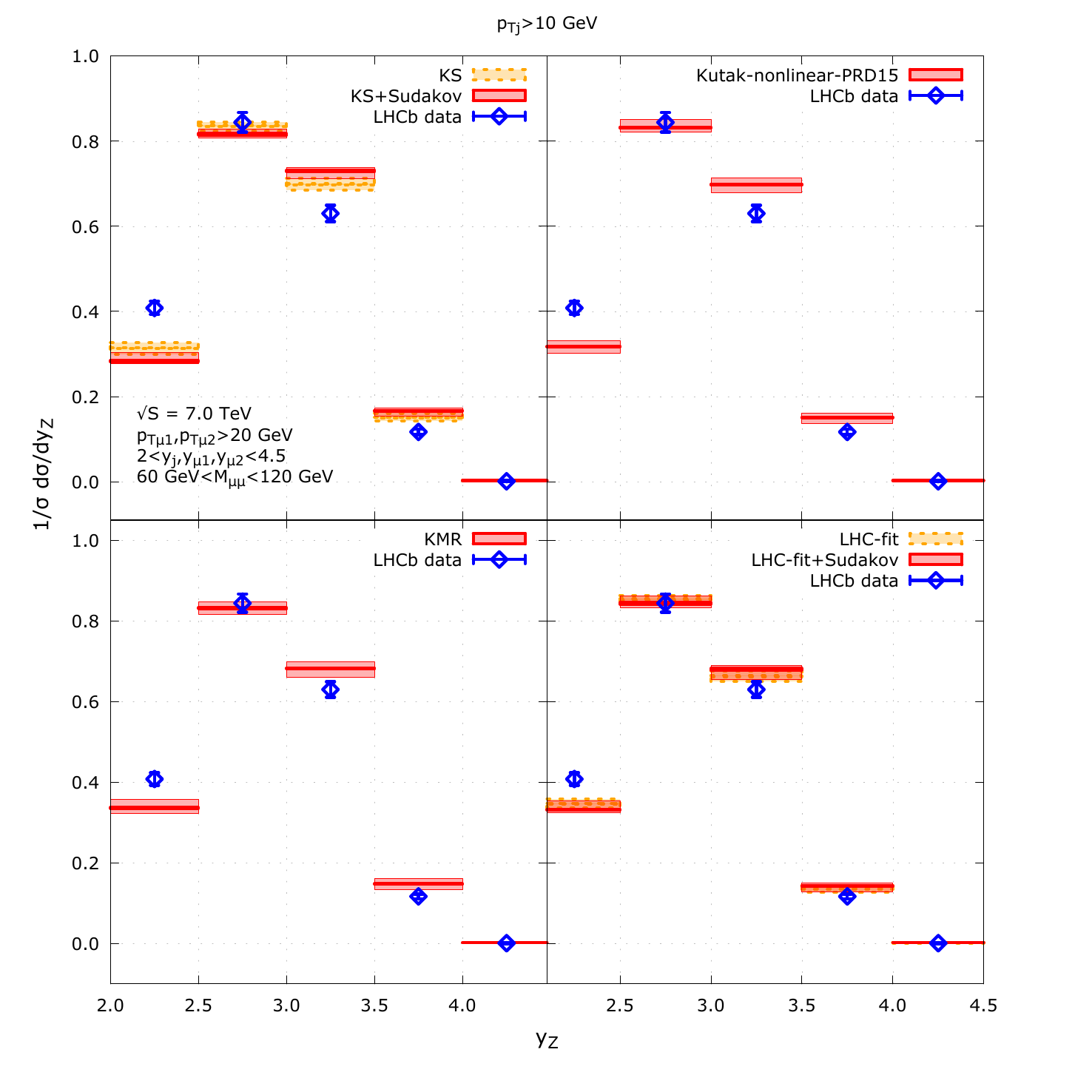}
\par\end{centering}

\caption{\small Rapidity spectrum of $Z_{0}$-boson for $p_{T\, j}>10\,\mathrm{GeV}$
normalized to the total cross section.\label{fig:yZ10}}
\end{figure}

\begin{figure}
\begin{centering}
\includegraphics{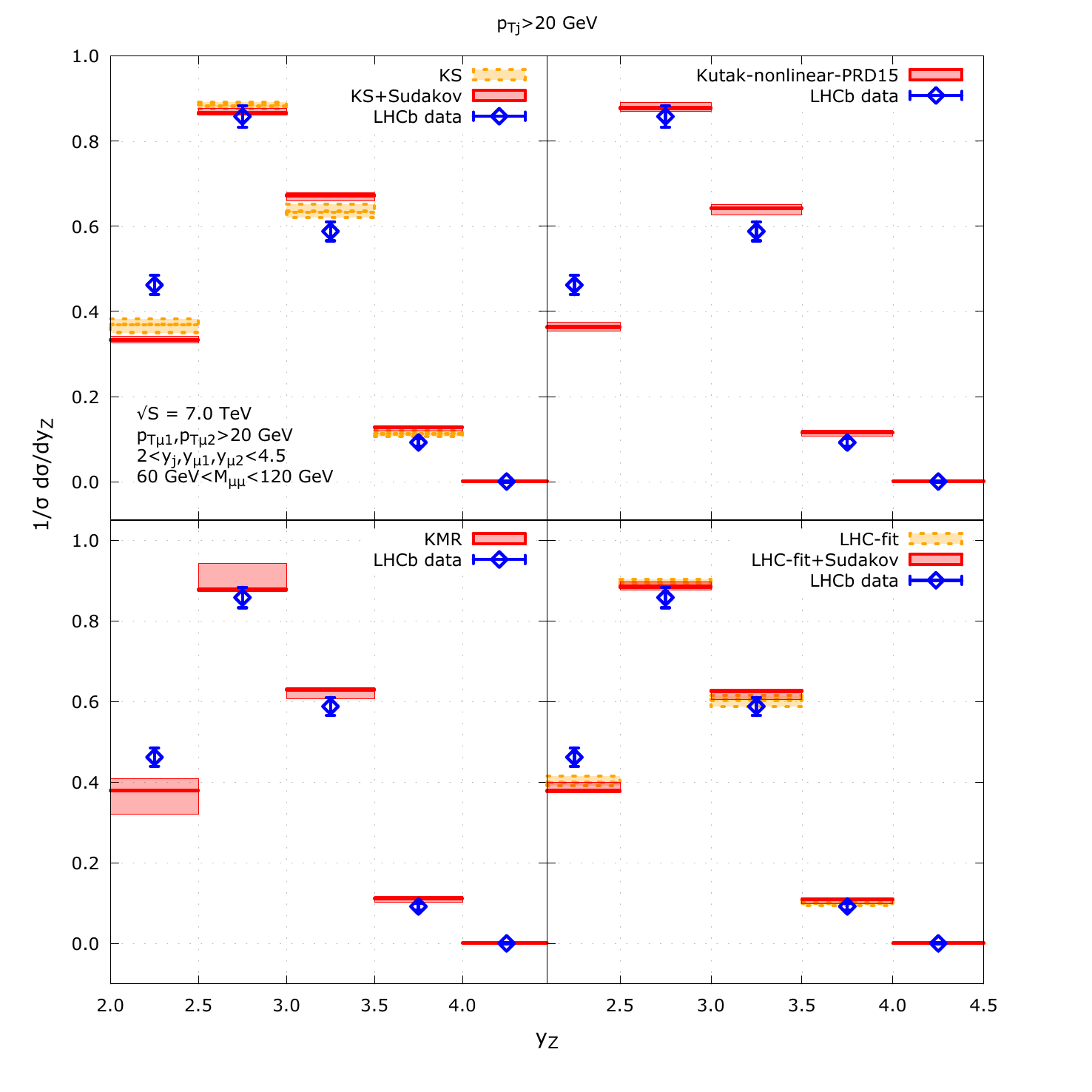}
\par\end{centering}

\caption{\small Rapidity spectrum of $Z_{0}$-boson for $p_{T\, j}>20\,\mathrm{GeV}$
normalized to the total cross section.\label{fig:yZ20}}
\end{figure}

\begin{figure}
\begin{centering}
\includegraphics{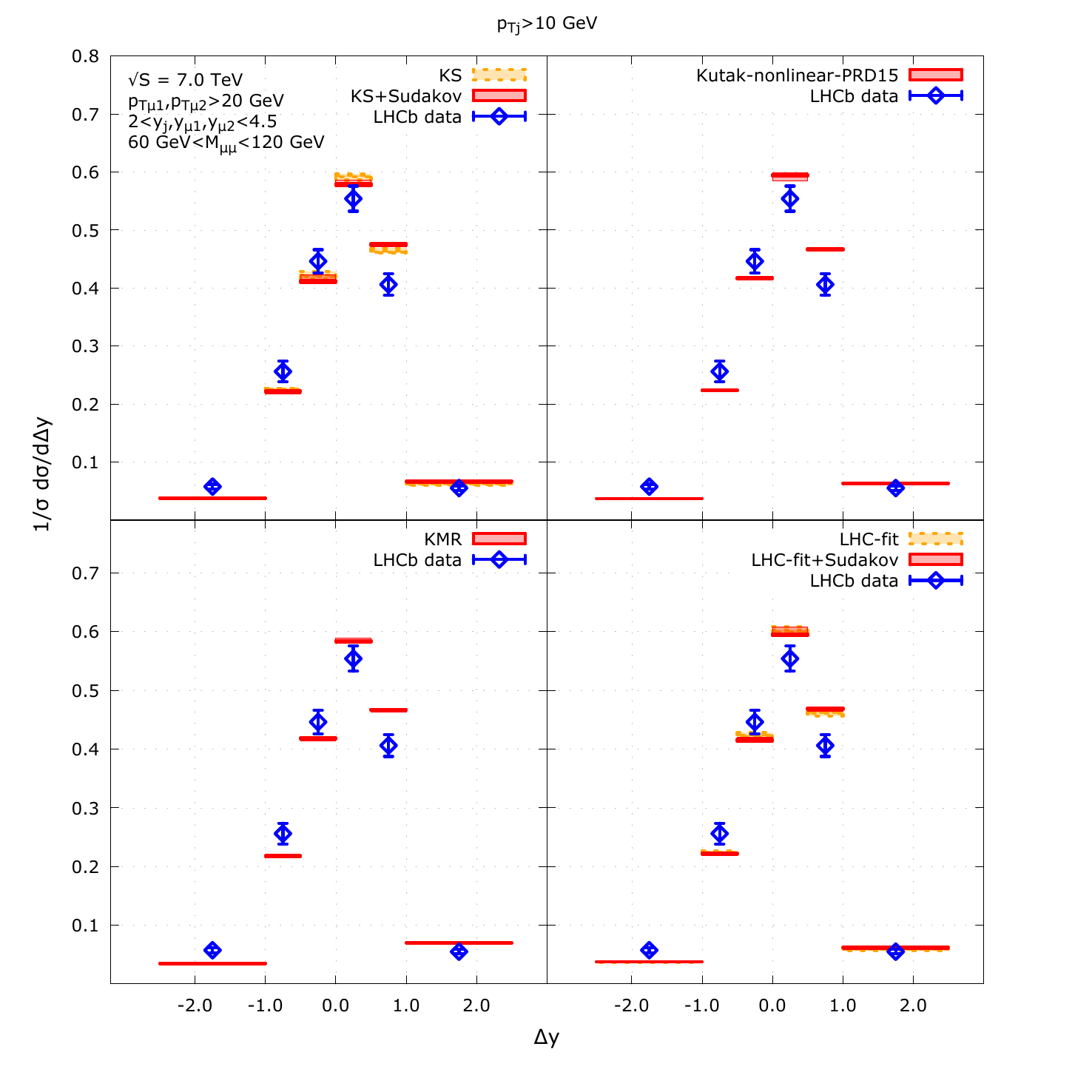}
\par\end{centering}

\caption{\small Differential cross section as a function of rapidity separation between
the $Z_{0}$-boson and the jet for $p_{T\, j}>10\,\mathrm{GeV}$ normalized
to the total cross section.\label{fig:dy10}}
\end{figure}

\begin{figure}
\begin{centering}
\includegraphics{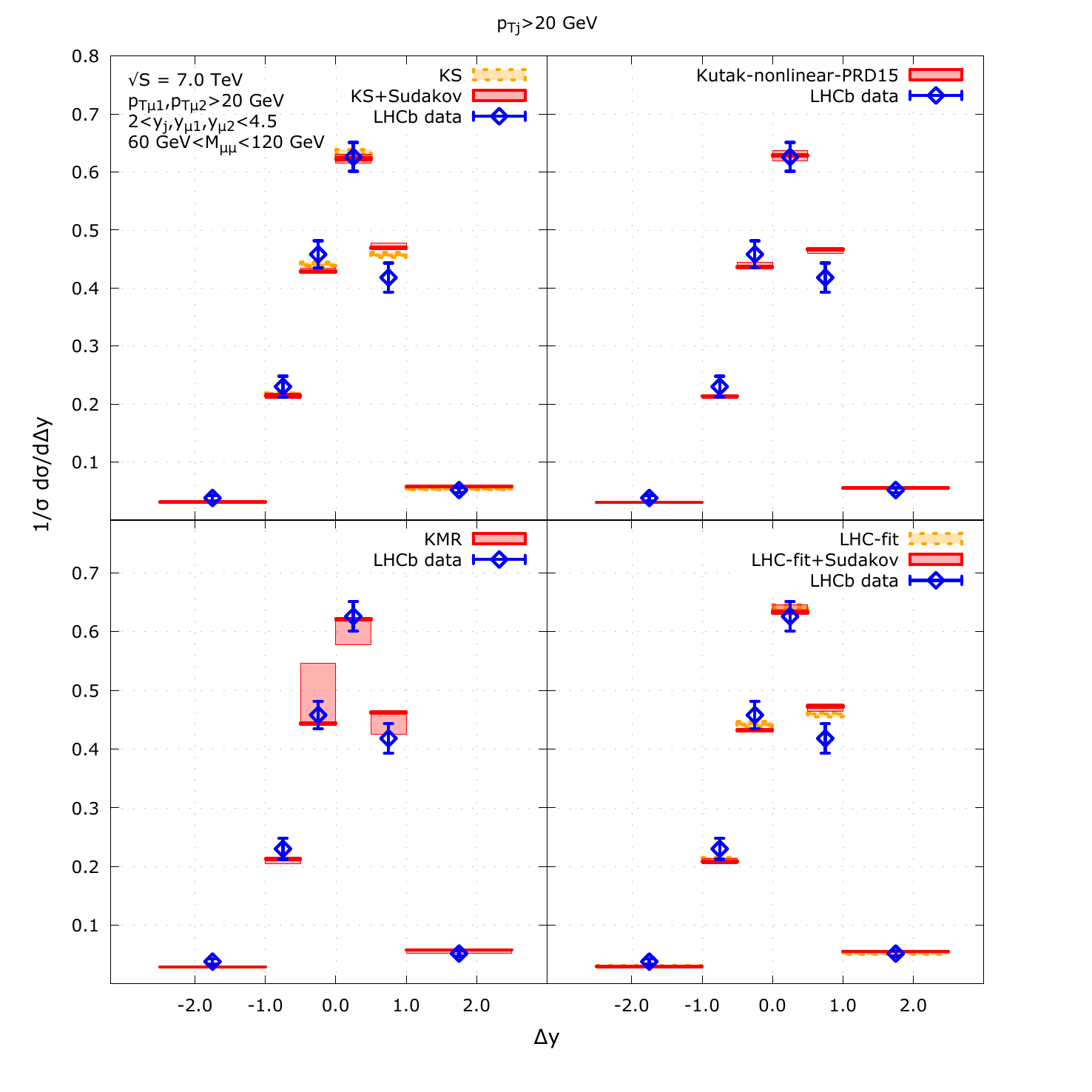}
\par\end{centering}

\caption{\small Differential cross section as a function of rapidity separation between
the $Z_{0}$-boson and the jet for $p_{T\, j}>20\,\mathrm{GeV}$ normalized
to the total cross section.\label{fig:dy20}}
\end{figure}

\begin{figure}
\begin{centering}
\includegraphics{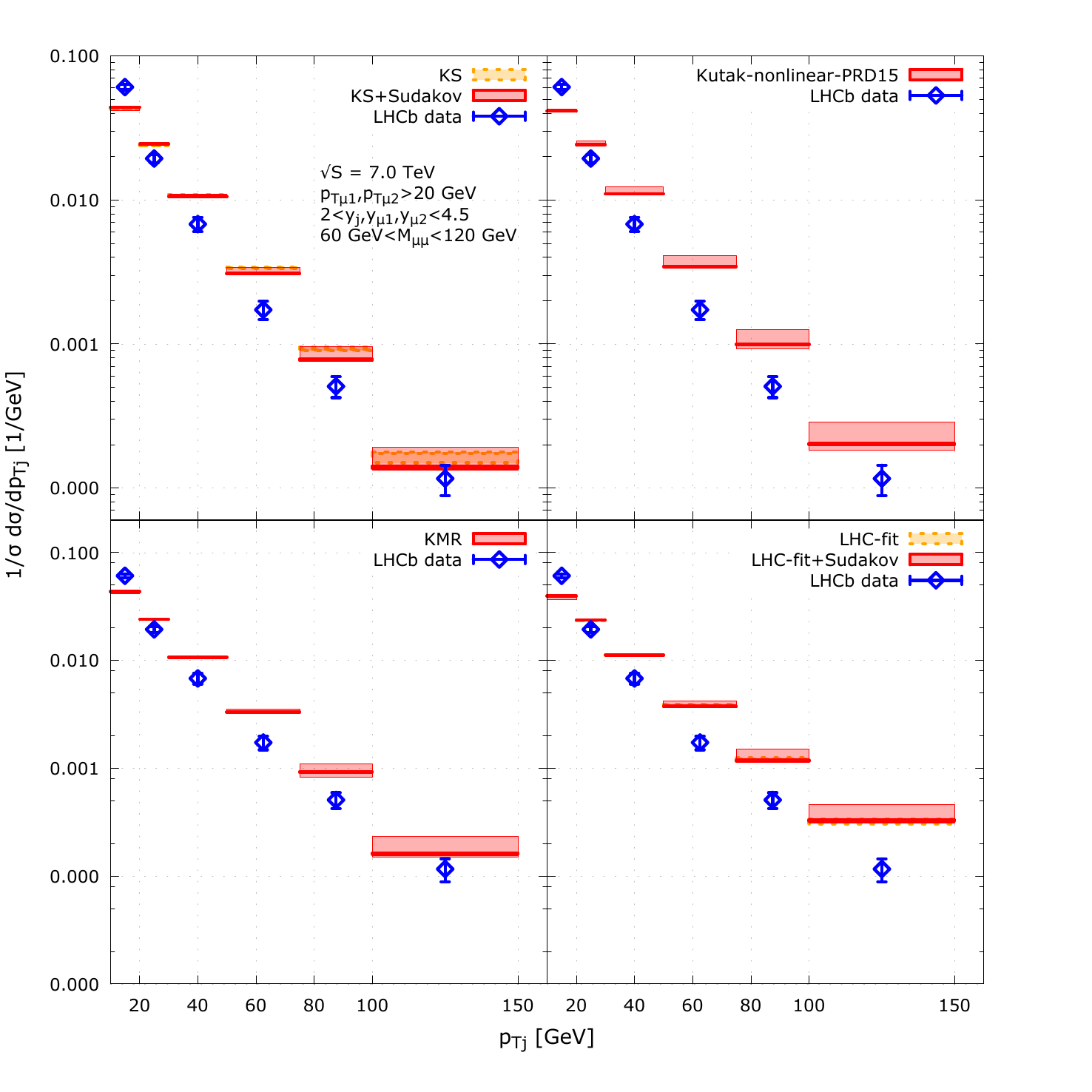}
\par\end{centering}

\caption{\small Transverse momentum spectrum of the jet normalized to the total cross
section.\label{fig:pTj}}
\end{figure}

For the numerical computation we use the kinematic cuts as described in the previous section, i.e. the ones used in
\cite{Aaij:2013nxa}. We use standard values for the electroweak
parameters: electroweak coupling $g_{\mathrm{ew}}=0.308$, $Z_{0}$
boson mass $M_{Z}=91.2\,\mathrm{GeV}$ and width $\Gamma_{Z}=2.495\,\mathrm{GeV}$.
We work with the 5 flavour scheme and use the CTEQ10NLO set \cite{Lai:2010vv} for the collinear
PDFs. For the unintegrated gluon distributions we consider the following models (in brackets with
bold font we give our abbreviations for the models):
\begin{itemize}
\item BFKL evolution with kinematic constraint, running strong coupling constant, DGLAP effects including the contribution from the quark sea \cite{Kwiecinski:1997ee}
and supplemented with the nonlinear term \cite{Kutak:2004ym}; the
initial condition has been fitted to HERA data \cite{Kutak:2012rf}
(\ugdKS)
\item hard scale dependent \ugdKS; the hard scale dependence is achieved by
implementing the Sudakov form factor on the generated events in such
a way that the total cross section remains unchanged \cite{vanHameren:2014ala}
(\ugdKSSud)
\item similar to above, but the Sudakov form factor is implemented to the
\ugdKS density in such a way that the integrated gluon density remains
unchanged \cite{Kutak:2014wga} (\ugdKSmu)
\item the simplified Kimber-Martin-Ryskin model \cite{Kimber:2001sc}
applied to MRSTW08 PDFs; this model gives unintegrated gluon distribution that is hard scale dependent
(\ugdKMR)
\item BFKL evolution with kinematic constraint, running strong coupling constant and the DGLAP contribution coming only from gluons 
and fitted to the LHC jet data \cite{Kotko:2015ksa} (\ugdLHC)
\item hard scale dependent LHC fit (\ugdLHCSud)
\end{itemize}
The hard scale was chosen to be the average of the three large scales appearing in the problem: the mass of the $Z_0$ boson, its $p_T$ and the $p_T$ of the jet. We present the results for numerical simulations in Figs.~\ref{fig:dphi10}-\ref{fig:pTj}
and compare them with LHCb data. 
Figs.~\ref{fig:dphi10},\ref{fig:dphi20} present normalized azimuthal decorrelations for $p_{Tj}>10\,\mathrm{GeV}$ and $p_{Tj}>20\,\mathrm{GeV}$ respectively (the azimuthal decorrelation is defined as the differential cross section in 
the difference of azimuthal angles of the reconstructed $Z_0$ boson and the leading jet). We present also results for normalized  differential cross sections in the transverse momentum $p_{T Z}$ of $Z_0$ boson (Figs.~\ref{fig:pTZ10},\ref{fig:pTZ20}), in rapidity $y_{Z}$ of $Z_0$  (Figs.~\ref{fig:yZ10},\ref{fig:yZ20}), in the transverse momentum $p_{Tj}$ of the jet (Fig.~\ref{fig:pTj}), and -- finally-- in the rapidity separation between the $Z_0$-boson and the jet $\Delta y$ (Figs.~\ref{fig:dy10},\ref{fig:dy20}). The shaded boxes in Figs.~\ref{fig:dphi10}-\ref{fig:pTj} represent the theoretical
uncertainty obtained by varying the hard scale by a factor of two.
The differential cross sections are normalized to the total cross section, as in \cite{Aaij:2013nxa}.
 We list the total cross sections for different models in Table~\ref{table1}.

\section{Discussion and conclusions}
\label{sec:discussion}

\begin{table}
\begin{doublespace}
\begin{centering}
\begin{tabular}{|c|c|c|}
\hline 
unintegrated gluon distribution model & $p_{T\, j}>10\,\mathrm{GeV}$ & $p_{T\, j}>20\,\mathrm{GeV}$\tabularnewline
\hline 
\hline 
\ugdKS & $4.1_{-0.7}^{+1.0}\,\mathrm{pb}$ & $2.3_{-0.4}^{+0.6}\,\mathrm{pb}$\tabularnewline
\hline 
\ugdKSmu & $4.1_{-1.0}^{+1.0}\,\mathrm{pb}$ & $2.4_{-0.5}^{+0.6}\,\mathrm{pb}$\tabularnewline
\hline 
\ugdKMR & $5.8_{-1.3}^{+1.5}\,\mathrm{pb}$ & $3.3_{-0.6}^{+0.9}\,\mathrm{pb}$\tabularnewline
\hline 
\ugdLHC %
 & $3.4_{-0.6}^{+0.8}\,\mathrm{pb}$ & $2.0_{-0.4}^{+0.5}\,\mathrm{pb}$\tabularnewline
\hline 
\hline 
LHCb data & $16.0\pm1.4\,\mathrm{pb}$ & $6.3\pm0.5\,\mathrm{pb}$\tabularnewline
\hline 
\end{tabular}
\par\end{centering}
\end{doublespace}

\caption{\label{table1}\small Total cross sections obtained from different models for the unintegrated
gluon density. The model uncertainties are defined through variations of the hard scale. For the data the total uncertainty is estimated as an
average square error from statistical, systematic and luminosity errors
as given in~\cite{Aaij:2013nxa}. We do not include models \ugdKSSud\ and \ugdLHCSud\ as the Sudakov-type resummation applied there does not change the total cross section.}

\end{table}

Let us first discuss the normalized differential cross sections. We can conclude that the azimuthal decorrelations are described  
reasonably well for both jet $p_T$ cuts for most of the models. It was however essential for this observable to include the hard scale dependence in unintegrated gluon distributions. As this observable is the most sensitive one for the small $x$ effects this underlines the importance of the resummation of the logarithms $\mathrm{ln}(\mu/k_T)$, where $\mu$ is the hard scale provided by the hard process and $k_T$ is the transverse momentum of the gluon in unintegrated gluon distributions, on the top of the small $x$ logarithms. The effect of this Sudakov-type resummation was also important for the transverse momentum spectrum of $Z_0$ boson. For the other observables it has had much less impact.

Since our calculations are not interfaced with any sort of final state parton shower we expected rather rough description of transverse momenta spectra. Our actual study shows however that the situation is relatively good for the spectrum of $Z_0$ boson (except very low transverse momentum), Figs.~\ref{fig:pTZ10}-\ref{fig:pTZ20}, and indeed fails for the jet spectrum, Fig.~\ref{fig:pTj}. This can be attributed to the fact that all unintegrated gluon distributions we have used contain contribution from pieces of splitting functions  subleading at low $x$ (see \cite{Kotko:2015ksa} for an analysis of the impact of this correction on jet $p_T$ spectra). This correction seems to be enough for the colorless final state while the color rescattering for the final state jet is evidently missing. Also, the next-to-leading correction in the hard process is necessary to improve the description of the transverse momentum spectra. The LHC-jet-motivated unintegrated gluon densities, which were actually fitted to the LHC jet transverse momentum spectra behave similar to the other gluon densities. This gives one more clue that the improvement in the hard process in terms of higher order corrections or/and a resummation is necessary. 

For $p_{Tj}>20\,\mathrm{GeV}$ the  \ugdKSmu\ and the \ugdKMR\ models overestimate the normalized transverse momentum spectrum of the $Z_0$ boson. These models implement a hard scale dependence but the spectra they give are almost the same as from the \ugdKS model, which is hard-scale-independent (the difference is however in the uncertainty which is much bigger for the hard-scale-dependent models). This means that the Sudakov-type resummation in the former models has no effect for this observable with such large $p_T$ cut. On the contrary, the Sudakov-type resummation with unitarity constraint (in the sense of preserving the total cross section) applied to the \ugdKS\ or \ugdLHC\ improves the results. 

Interestingly, the unintegrated gluon distribution which comes from the fit to the LHC jet data does not perform better that the distributions obtained from the inclusive DIS data. This suggests that the effects of factorization breaking due to the lack of universality is rather weak at the phenomenological level. As already mentioned above, this suggests also that one needs higher order corrections in the hard process. Indeed, as observed in \cite{Kotko:2015ksa} the $p_T$ spectra cannot be described by improving the evolution of unintegrated gluon distribution itself. One should mention, that the evolution equations that were fitted to the LHC data did not include a contribution from the sea quarks on the dense hadron side, whereas this contribution is present in the other approaches. It is however unlikely that this can improve the situation. 

Present calculations did not take into account the situation where the off-shell initial-state parton is a quark or an anti-quark. We expect this contribution to be small, as at small $x$ gluons dominate significantly. On the other hand, the process $\bar{q}q\rightarrow Z_{0}\left(\rightarrow\mu^{+}\mu^{-}\right)+g$ gives an important contribution in the collinear approach. Since in reality the probed values of $x$ are not extremely small, such process might be important in the High Energy Factorization. Practical applications require however a set of unintegrated quark distributions, consistent with the unintegrated gluon distributions. Inclusion of those is left for future studies.

As seen from the Table~\ref{table1} our calculations strongly underestimate the total cross section as compared to the data. This should be probably attributed to the hard multi-parton interaction (MPI) effects which are not taken into account in the Hybrid High Energy Factorization. This statement is supported by the observation that the description of the azimuthal decorrelations is very good up to the normalization, which, when corrected, will only shift the data eventually forming  a ``pedestal''. It is also known that MPIs indeed contribute only a pedestal to the azimuthal decorrelations (see e.g. \cite{Strikman:2010bg,Stasto:2011ru}), as the two partons coming from independent scatterings are completely decorrelated (to leading order) and thus the azimuthal decorrelation distribution is flat. The subject of MPIs in the High Energy Factorization is however rather complicated and needs a separate study. In principle, on the dense hadron side (i.e. for the one probed at small $x$), the soft MPIs can be partially taken into account by means of the nonlinear term in the evolution equation, as for example in the \ugdKS\ unintegrated gluon density or its extensions. Therefore, one has to be careful not to make a double counting when using some phenomenological models by an inclusion of double hard scattering mechanism. Actually the problem extends beyond the saturation regime as the High Energy Factorization contributes to higher twists as well as MPIs, at least for certain observables \cite{Diehl:2011yj}. These subjects are however beyond the scope of the present work.

\section*{Acknowledgments}
We acknowledge useful correspondence and discussions with Albert Bursche, and we thank Rafa\l Maciu\l{}a for providing us the necessary grid for the KMR pdf.
The work of K.K. has been supported by Narodowe Centrum Nauki with Sonata Bis grant DEC-2013/10/E/ST2/00656.
P.K. acknowledges the support of the grants DE-SC-0002145 and DE-FG02-93ER40771.


\providecommand{\href}[2]{#2}\begingroup\raggedright\endgroup

\end{document}